\pdfoutput=1

\documentclass[11pt]{article}

\usepackage{coling}

\usepackage{times}
\usepackage{latexsym}

\usepackage[T1]{fontenc}

\usepackage[utf8]{inputenc}

\usepackage{microtype}

\usepackage{amsmath}
\usepackage{booktabs}
\usepackage{cleveref}
\usepackage{graphicx}
\usepackage{multirow}
\usepackage{enumitem}
\usepackage{hyperref}
\usepackage{subcaption}

\usepackage{ai-usage-card}
\usepackage{annotation}

\title{
\vspace*{-0.5in}
{\raggedright\small COLING 2025\\[0.25in]}
Citation Amnesia:\\On The Recency Bias of NLP and Other Academic Fields
}

\author{Jan Philip Wahle${^\Omega}$${^\Theta}$\textsuperscript{*}, Terry Ruas${^\Omega}$, Mohamed Abdalla$^{\Phi}$, Bela Gipp${^\Omega}$, Saif M. Mohammad$^{\Theta}$\\
$^{\Omega}$University of Göttingen, Germany\\
$^{\Phi}$University of Alberta, Canada\\
$^{\Theta}$National Research Council, Canada\\
\textsuperscript{*}\texttt{wahle@uni-goettingen.de}\\}

\begin{document}
\maketitle
\AddAnnotationRef

\begin{abstract}
This study examines the tendency to cite older work across 20 fields of study over 43 years (1980--2023). We put NLP's propensity to cite older work in the context of these 20 other fields to analyze whether NLP shows similar temporal citation patterns to them over time or whether differences can be observed. Our analysis, based on a dataset of $\approx$240 million papers, reveals a broader scientific trend: many fields have markedly declined in citing older works (e.g., psychology, computer science).
The trend is strongest in NLP and ML research (-12.8\% and -5.5\% in citation age from previous peaks). Our results suggest that citing more recent works is not directly driven by the growth in publication rates (-3.4\% across fields; -5.2\% in humanities; -5.5\% in formal sciences) --- even when controlling for an increase in the volume of papers. Our findings raise questions about the scientific community's engagement with past literature, particularly for NLP, and the potential consequences of neglecting older but relevant research. The data and a demo showcasing our results are publicly available.\footnote{\href{https://github.com/jpwahle/coling2025-citation-age}{https://github.com/jpwahle/coling2025-citation-age}}
\end{abstract}

\section{Introduction}

Innovations arise on the backs of past ideas and from the cross-fertilization of ideas. 
Researchers discuss related work from various fields of study to confirm or reject earlier findings, to compare and situate the proposed work, and, ultimately, to build on previous ideas.
Citations\footnote{We acknowledge that there are other equally important proxies such as h-index, engagement, etc.} 
are a primary mechanism to acknowledge influence and guide readers through related ideas. 
Analyzing citation patterns offers insight into the values of a field, revealing what is considered important, what may be overlooked, and where it is headed. %

\begin{figure}[t]
    \centering
    \includegraphics[width=0.9\columnwidth]{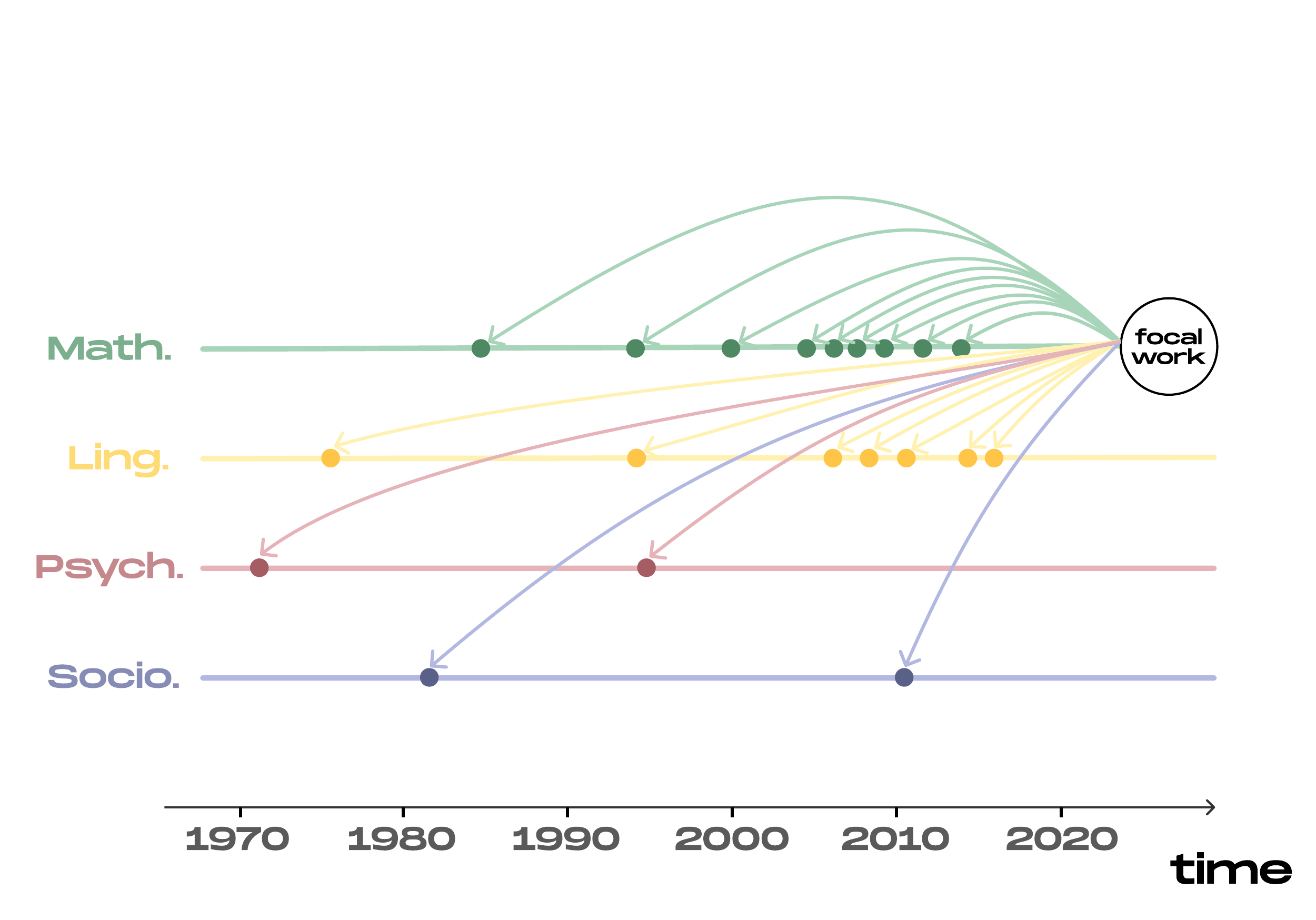}
    \caption{Scientific works cite others across fields and time. A focal work may cite works from its own or other fields and in varying degrees from the past.}
    \label{fig:teaser}
    
\end{figure}

Responsible research should arguably engage with a broad set of literature, spanning from various fields and periods \cite{burget2017definitions,wahle-etal-2023-cite}. 
\Cref{fig:teaser} illustrates a focal work and how it cites works from various other fields across different points in time --- tracing how these citation patterns change is necessary to foster robust and inclusive scientific discourse \cite{bollmann-elliott-2020-forgetting,singh-etal-2023-forgotten}. 

Of particular interest is the tendency to not cite enough relevant good work from the past (more than a few years old) --- \textit{citation amnesia} \cite{garfield1980citation, garfield1982more, singh-etal-2023-forgotten}.
This trend can stem from various factors, including the deliberate omission of known work, unintentional forgetfulness, or simply a lack of awareness about pertinent research, especially when it originates from fields different from the author's field.
Determining how much `relevant' or `good' old work is forgotten requires expert researcher judgment and is subjective, making empirical measurements of citation amnesia challenging \cite{singh-etal-2023-forgotten}. 
However, we can measure the collective tendency of a field to cite older work (from within its field or from other fields). 
A dramatic change in our tendency to cite older work should encourage us to reflect on whether we are putting enough effort into reading older papers.
We are not calling for citing works just because they are old but to reflect on the broad trends of how much a field cites older work. %

As researchers, we play an active role in how much older work is forgotten. 
We are free to choose which literature to engage with. 
Forgetting some old works can be helpful, as it makes space for new ideas.
However, too much forgetting can lead to an unnecessary reinvention of concepts and methods. 
We want to avoid neglecting older works by lack of engagement in favor of consciously deciding specific older works may not be relevant to us. %

Studying temporal citation patterns is vital for any field, but we argue NLP deserves specific attention because its interdisciplinary nature inherently influences various other fields, such as linguistics, psychology, and computer science (CS). NLP advancements like large language models have captured the world's imagination and are poised to influence societies and industries substantially.
Recent studies have focused mainly on temporal citation patterns within NLP and show a marked decline in citing old works starting in $\approx$2015.
However, they are not concerned with the citation dynamics of other fields and the temporal cross-field interaction between NLP and other fields \cite{bollmann-elliott-2020-forgetting,singh-etal-2023-forgotten}.
These analyses, therefore, take one specific vertical slice of \Cref{fig:teaser}. 
We do not know whether these trends can be observed for other fields too, specifically those that NLP interacts with frequently.

This study systematically examines temporal citation patterns across NLP and 20 other fields over 43 years. 
We quantitatively measure temporal citation patterns between NLP and other fields. 
We answer eight research questions (\Cref{sec:experiments}) grouped into four broad questions:

\begin{enumerate}[itemsep=1pt]
    \item How much do papers of various fields cite relatively old work, and how does that change over time?
    \item Which fields cite older works more? Which fields cite older works less?
    \item How does NLP's tendency to cite old works compare to other fields? How far back in time do NLP papers cite works from other fields?
    \item Does the temporal distribution of citations correlate with cross-field engagement (another important facet of responsible research)?
\end{enumerate}

The primary audience of this work is NLP researchers. NLP is a multidisciplinary field, and its applications have a broad social impact. Innovations in recent years have greatly increased the reach of NLP to the masses worldwide. The importance of responsible and sustainable research practices has never been higher. By situating NLP's tendency to cite older works within broader scientific trends across fields, we gain insights into how our field interacts with its own and other fields' intellectual history.

Every field needs to examine itself critically. The conferences and journals of a field are the best venues for such an examination; publishing self-critical work shows the world that we do not hide away from changing trends in our field and are working towards improving things if necessary. A long history of self-analytical work in NLP shows the importance of self-reflection \cite{radev-etal-2009-acl, gupta2011analyzing, vogel-jurafsky-2012-said, anderson-etal-2012-towards, gildea-etal-2018-acl, schluter-2018-glass, abdalla-etal-2023-elephant, wahle-etal-2023-cite}. Further, the ACL 2020 theme track ``Taking Stock of Where We've Been and Where We're Going'' and several workshops (e.g., SDProc\footnote{\href{https://sdproc.org/2022/}{https://sdproc.org/2022/}}, SciNLP\footnote{\href{https://scinlp.org}{https://scinlp.org}}) underline this.

Researchers from other fields can also benefit from our examination as we provide the results for each of the 23 fields and the source code to reproduce analyses even for individual subfields. We further provide recommendations on engagement with literature from the past and across fields.

\section{Related Work}

Scientometrics, and specifically the study of citation patterns, has garnered marked attention, focusing on various dimensions such as field of study \cite{costas2009scaling}, author affiliation \cite{sin2011international,abdalla-etal-2023-elephant}, paper length \cite{falagas2013impact}, publication venue \cite{callaham2002journal, wahle-etal-2022-d3}, paper quality \cite{buela2010analysis}, publication language \cite{lira2013influence},
geographic location \cite{rungta-etal-2022-geographic},
gender \cite{mohammad-2020-gender,chatterjee2021gender,abdalla2023ethnicity},
self-citation \cite{della2008multi}, industry presence \cite{abdalla-etal-2023-elephant}, plagiarism \cite{gipp2011citation, wahle-etal-2022-large}, paraphrase \cite{wahle-etal-2023-paraphrase},
and author reputation \cite{castillo2007estimating, petersen2014reputation}.

An area of recent particular interest is the temporal aspect of citations, specifically citation amnesia. 
The term `citation amnesia' was already used by \citet{garfield1980citation} in the early 80s to describe the tendency not to cite potentially relevant related works and was picked up later by others such as \citet{rilling1996mystery,maes2015review,singh-etal-2023-forgotten}. 

Work by \citet{verstak2014shoulders} analyzed scholarly articles published between 1990 and 2013, revealing an increasing trend in citing older papers, which they attributed to easier access to scientific literature online. 
In CS, there was a 39\% increase from 1990 to 2013 in citing papers over ten years old. 
\citet{parolo2015attention} extended this analysis to fields like clinical medicine, physics, and chemistry, observing that the peak citation period of papers is followed by an exponential decay, with this decay rate increasing in more recent publications. 

\citet{bollmann-elliott-2020-forgetting} examined the recency bias in citations in NLP, showing that papers from 2010 to 2014 have cited, on average, more older papers when compared to those from 2017 to 2019. 
\citet{singh-etal-2023-forgotten} extended this investigation to a broader range of 70k+ NLP papers between 1965--2021, showing that NLP articles from 1990--2014 were increasingly citing older papers. However, starting in 2015, an abrupt drop in old citations uncovers NLP's tendency toward recent publications. 
Contemporary to our work, \cite{nguyen2024there} have analyzed citation amnesia of various fields of study in arXiv. This shows that traditional fields, such as math or physics, have not experienced a recency bias in their citations.

Our research expands upon previous findings by analyzing a dataset covering a broader range of 20 high-fields (e.g., math, psychology) and three subfields of CS (i.e., NLP, ML, and AI) and a longer period of 43 years.
In addition to \citet{singh-etal-2023-forgotten}, who documented a shift towards citation amnesia within NLP, our analysis across 20 fields provides insights into broader trends of citation amnesia as well as the temporal citation interactions between NLP and other fields. 
Tracing temporal patterns for cross-field citations is inspired by \citet{wahle-etal-2023-cite}'s findings on the declining cross-field engagement within NLP, which did not look at temporal citation patterns. 
\citet{nguyen2024there}'s observation of citation amnesia across different quantitative fields in arXiv is complemented by our work, which situates these patterns within a larger set of both quantitative and non-quantitative fields with a larger corpus across a longer period from 1990 to 2023. Going beyond how overall temporal citation patterns have changed, our work goes into other novel research questions, notably around intra-field and inter-field citations (Q1, Q2, Q3, Q4), citation ages of NLP to specific other fields (Q5, Q6), and whether NLP cites the same old works over time (Q7).

\section{Data}

Central to a study examining citation age across various fields of study is a dataset that includes the field of study of a paper and its publication year. 
It should be noted that a paper can be associated with multiple fields in varying degrees, making it challenging for both humans and automated systems to assign these labels and scores accurately (e.g., a paper about the use of AI in medicine). 
Additionally, acquiring a comprehensive collection of papers for each field is challenging, as defining the boundaries of a field itself is complex. 
Despite these issues, at an aggregate level, important inferences about the citation dynamics of a field can be drawn.

We derive data from OpenAlex \cite{priem2022openalex}, a repository with $\approx$240m papers and $\approx$280b citations under the CC0\footnote{\href{https://creativecommons.org/publicdomain/zero/1.0/}{https://creativecommons.org/publicdomain/zero/1.0/}} license (for exact numbers, see \Cref{tab:paper_counts_by_field} in \Cref{sec:appendix}). 
The dataset contains 20 high-level fields, such as psychology, math, and CS, as well as their first-level subfields, such as algorithms and databases (for CS), and second-level subfields, such as greedy methods and linear programming (for algorithms). 
NLP, ML, and AI are direct children of CS, although an NLP paper can be part of multiple fields in the dataset (e.g., linguistics, psychology, and CS). 

We sample 1\% of papers per field\footnote{But at least 30,000 examples.} to reduce computational costs and report results with 95\% confidence intervals (for more details on the number of papers, see \Cref{ap:data}). 
The source code used in processing our data and conducting experiments is available on GitHub\footnote{\href{https://github.com/jpwahle/coling2025-citation-age}{https://github.com/jpwahle/coling2025-citation-age}}

\section{Analysis}
\label{sec:experiments}

We use the dataset described above to answer a series of questions about citation amnesia of NLP and various other fields.\\[3pt]
\noindent\textbf{Q1.} \textit{How far back in time do we go to cite papers? As in, what is the average age of cited papers? How does it differ across different fields?}
\noindent \textbf{Ans.} Following \citet{bollmann-elliott-2020-forgetting,singh-etal-2023-forgotten} for each paper in a field, we look at the citations to other papers and compute how far back in time the current paper is citing. When a paper $x$ cites a paper $y_i$, then the age of the citation (AoC) is the difference between the year of publication (YoP) of $x$ and $y_i$:

\begin{equation}
    \text{AoC}(x, y_i) = \text{YoP}(x) - \text{YoP}(y_i)
\end{equation}

We calculate the \textit{mean} AoC for each of the citations of a paper and average them:

\begin{equation}
    mAoC(x) = \frac{1}{N} \sum_i^N \text{AoC}(x, y_i)
\end{equation}

\noindent where $N$ refers to the number of papers cited by $x$.

For example, if a paper $x$ from 2020 cites two papers, one from 2010 and one from 2000, the $mAoC$ of paper $x$ is 15 years.

\noindent\textbf{Results.} \Cref{tab:mean_aoc_per_field} shows the mean $mAoC$ for all papers of a field for the 20 fields of study and for NLP and ML. Observe how NLP has the lowest mean $mAoC$ of 9.44, with ML following closely with a mean $mAoC$ of 9.63.

Unsurprisingly, history has the highest mean $mAoC$ of 14.90. Fields with a long history have high mean $mAoC$, too (philosophy: 11.69; sociology: 11.20; economics: 10.40). For example, western philosophy has its origins already in ancient Greece in the 6th century BCE with major figures like Socrates, Plato, and Aristotle. Physics has been studied since the Renaissance with the work of Copernicus, Galileo, Kepler, and Newton.

\textbf{Discussion.} Different fields have varying citation dynamics. Traditional fields, such as history or philosophy, intuitively cite older papers; younger fields, like CS or engineering, predominantly work on the edge of innovation and thus frequently cite more recent studies. Medicine is an outlier with roots in ancient history but a particularly low mean $mAoC$. Medicine has a long history but is also characterized by disruptions over time; many treatment methods have been innovated. Another confounding factor could be that many medical journals limit the number of references, making it more likely to cite recent studies over foundational works. 

Citation dynamics can also be subfield-dependent. History works concerned with ancient history tend to cite much older works than those concerned with modern history. Yet it is an open question whether NLP should have a (much) lower $mAoC$ than two fields that form its interdisciplinary intersections: CS and linguistics. Also, it is yet unclear whether NLP and other fields have always had such citation ages and how these citation trends have evolved over time, i.e., whether there exist a trend of increasing or declining citation age between NLP and other fields. \\\newline
\noindent\textbf{Q2.} \textit{How has the average age of citation evolved over time, and how does this evolution differ across various fields?}

\begin{table}[t]
\centering
\resizebox{\columnwidth}{!}{
    \begin{tabular}{lr}
      \toprule
      Field & $mAoC$ ± 95\% Conf. ($\uparrow$) \\ 
      \midrule
      NLP$^*$ & 9.44 ± 0.14 \\ 
      Medicine & 9.47 ± 0.13 \\ 
      Engineering & 9.53 ± 0.14 \\ 
      ML$^*$ & 9.63 ± 0.12 \\ 
      Business & 9.84 ± 0.15 \\ 
      Chemistry & 10.03 ± 0.13 \\ 
      Computer science & 10.14 ± 0.16 \\ 
      Biology & 10.14 ± 0.15 \\ 
      Materials science & 10.20 ± 0.14 \\ 
      Environmental science & 10.32 ± 0.14 \\ 
      Economics & 10.40 ± 0.16 \\ 
      Political science & 10.73 ± 0.18 \\ 
      Psychology & 10.70 ± 0.14 \\ 
      Physics & 10.75 ± 0.16 \\ 
      Sociology & 11.20 ± 0.17 \\ 
      Geography & 11.24 ± 0.21 \\ 
      Mathematics & 11.52 ± 0.16 \\ 
      Linguistics & 11.61 ± 0.19 \\ 
      Philosophy & 11.69 ± 0.18 \\ 
      Geology & 11.76 ± 0.20 \\ 
      Art & 13.06 ± 0.23 \\ 
      History & 14.90 ± 0.28 \\ 
    \bottomrule
    \end{tabular}
}
\caption{The $mAoC$ and confidence intervals for different fields of study are ordered by increasing $mAoC$. $^*$Subfields of CS.\vspace*{-4mm}}
\label{tab:mean_aoc_per_field}
\end{table}

\begin{figure*}[t]
    \centering
    \begin{subfigure}[b]{0.3\textwidth}
        \centering
        \includegraphics[width=\textwidth]{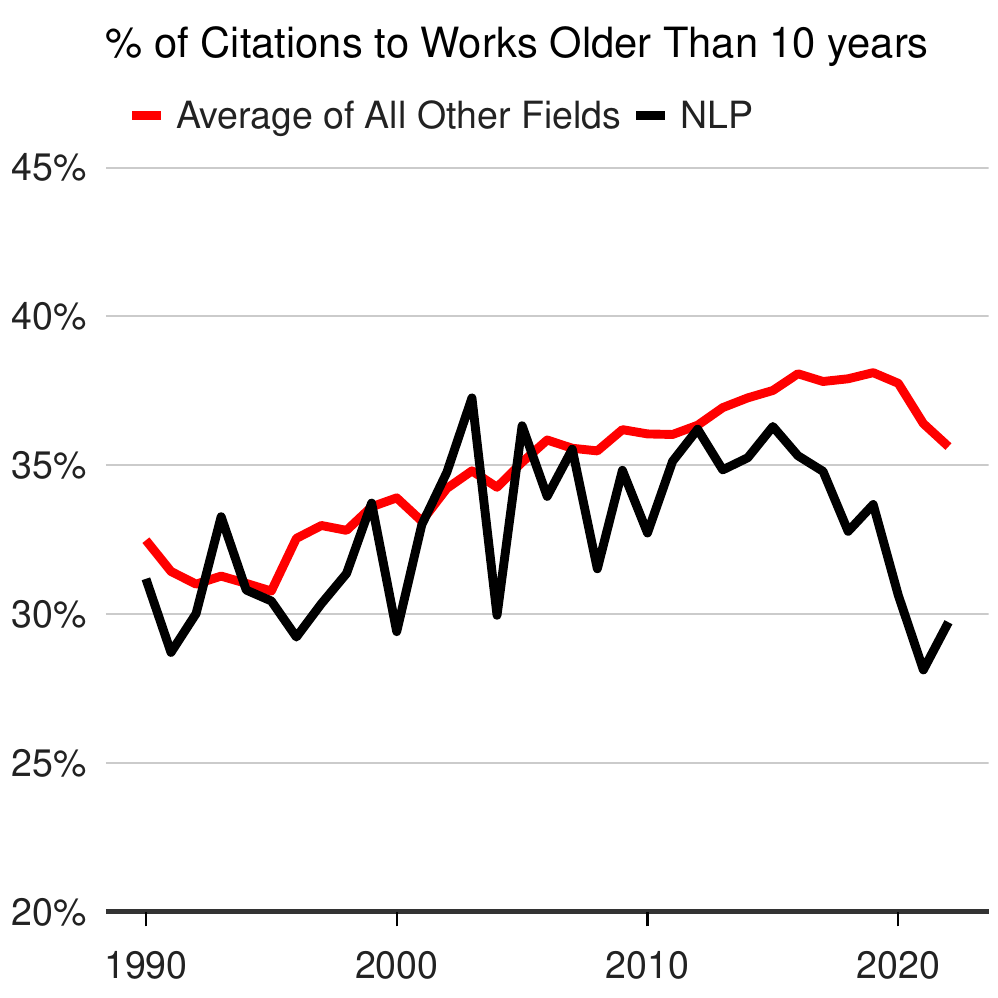}
        \caption{}
        \label{fig:perc_citations_to_older_works_nlp_other_fields}
    \end{subfigure}
    \hfill
    \begin{subfigure}[b]{0.3\textwidth}
        \centering
        \includegraphics[width=\textwidth]{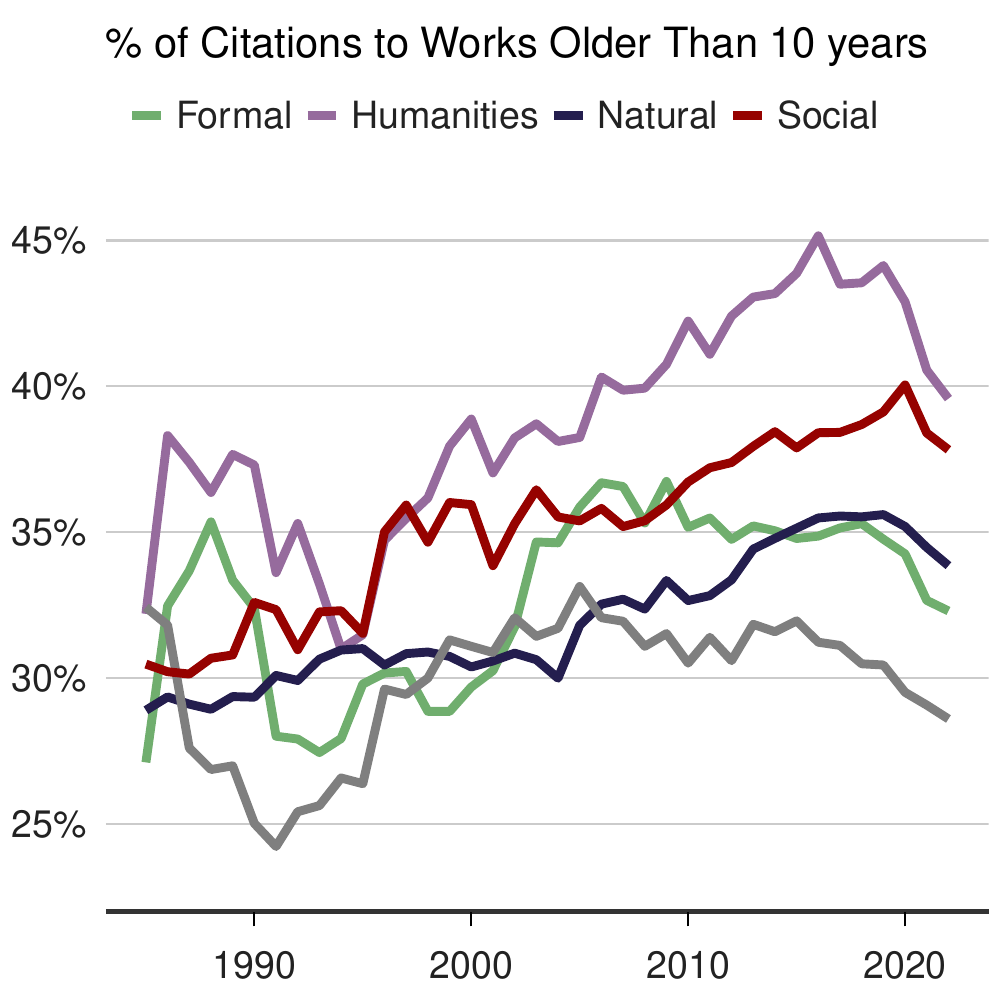}
        \caption{}
        \label{fig:perc_citations_to_older_works_field_groups}
    \end{subfigure}
    \hfill
    \begin{subfigure}[b]{0.3\textwidth}
        \centering
        \includegraphics[width=\textwidth]{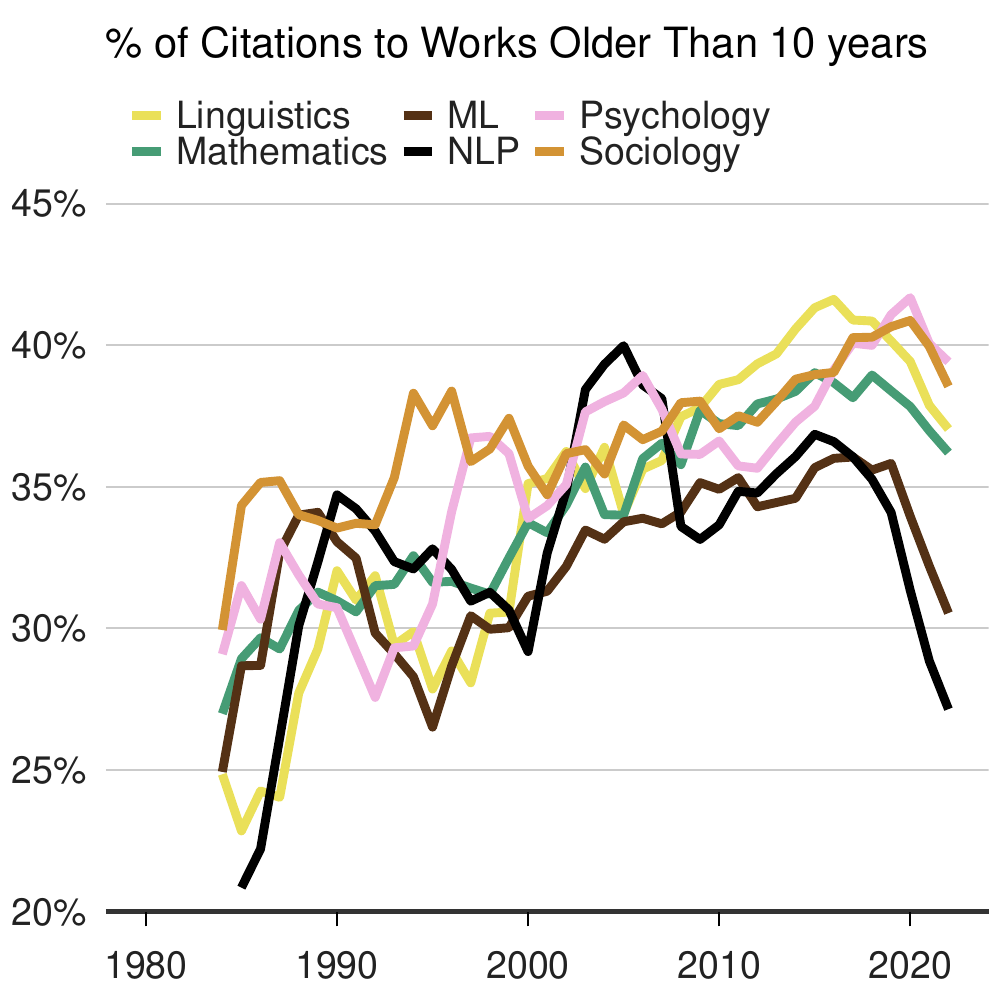}
        \caption{}
        \label{fig:perc_citations_to_older_works}
    \end{subfigure}
    \caption{The percentage of citations older than ten years for \textbf{(a)} NLP and the avg. of all 20 fields; \textbf{(b)} four field groups (top to bottom in 2023: humanities, social, natural, and formal sciences); \textbf{(c)} NLP, ML, and the top four cited fields by NLP (top to bottom in 2023: psychology, sociology, linguistics, math, ML, NLP).}
    \label{fig:three_subfigures}
\end{figure*}

\noindent\textbf{Ans.} We trace the percentage of citations to old papers (older than ten years) for the 20 fields. We also aggregate related fields into formal sciences, social sciences, natural sciences, and humanities to trace broader trends of change across fields.

\noindent\textbf{Results.} \Cref{fig:perc_citations_to_older_works_nlp_other_fields} shows the percentage of citations to works older than ten years. %
While NLP has increasingly cited older works from 1990 to 2015, it has seen a marked decline from all-time highs in 2015 (-12.8\%); other fields have also cited more older papers until 2019 but then saw a decline from their peaks (-2.2\%). 
\Cref{fig:perc_citations_to_older_works_field_groups} decomposes the average into four broad categories of fields of study, according to Wikipedia's categorization\footnote{\href{https://en.wikipedia.org/wiki/List_of_academic_fields}{https://en.wikipedia.org/wiki/List\_of\_academic\_fields}} of academic fields: formal sciences, natural sciences, social sciences, and humanities. The graph indicates a general trend across all fields towards citing fewer older works in recent years. 
Humanities have the highest percentage of citations to older works, peaking around 2015 before a stark decline. 
Social sciences also display a high and increasing percentage up to around 2018, before a noticeable decline. 
Natural sciences experienced a steady increase until around 2013, followed by a plateau and a slight decrease after 2019. 
Finally, the formal sciences (including CS) show the lowest percentage throughout, with a more variable trend line but an overall decline from a peak near 2010.

\Cref{fig:perc_citations_to_older_works} shows NLP, ML, and the four most cited fields by NLP. Both ML and NLP, %
have seen a stark relative decline in citations to older papers since 2015. NLP has declined by 12.8\% and ML by 5.5\% from their previous peaks. 
However, many fields have seen a marked relative decline in citations to older works between 2015 and 2020. 
Linguistics and math started to follow a downward trend in 2015 and 2017, respectively, with -4.6\% and -2.8\% from previous all-time highs. 
Psychology and sociology only recently started a downward trend in 2020 by a few percentage points.

\noindent\textbf{Discussion.} Contrary to \citet{nguyen2024there}, our results show a trend of reduced citations to older works across many fields. %
These newly uncovered trends reveal marked shifts, and there could be much more downward potential in this 'recession` before trends return to pre-2015 conditions. The reason behind this trend remains uncertain, but something affects how far back in time we cite.
Whatever the cause, it appears that we are at the beginning of a broader scientific phenomenon.\\[3pt]
\noindent\textbf{Q3.} \textit{How much does the volume of papers affect $mAoC$? Did the rate of increase in the number of cited papers grow substantially in 2015?}

\noindent\textbf{Ans.} This question is motivated by contemporary work suggesting that if there is a growing number of papers in a field (e.g., CS), then this field is more likely to cite recent papers than papers from 10 years ago \cite{nguyen2024there}. In the following, we investigate whether the volume impacts citation age as the growth in volume is not unique to any single field; many academic fields are experiencing growth in the number of published papers, whereas their trends to cite recent work show vital differences. 
The value of a paper does not necessarily diminish over time. 
Foundational theories and long-standing principles remain relevant, such as Newton's laws of motion; newer papers still build upon these established ideas. 
The algorithms of search engines, often used for literature research, also consider other factors than publication date, e.g., number of citations, publication venue \cite{beel2009google,ValenzuelaEscarcega2015IdentifyingMC}.

We introduce Volume-Adjusted Average Citation Age (VACA), a metric that normalizes the $mAoC$ of a field by the number of papers. By controlling citation age with the number of papers in that year, we can account for exponential changes in volume and whether they impact $mAoC$. The metric can be computed as:
\begin{equation}
    \textnormal{VACA} = \frac{mAoC}{V_{norm}}
\end{equation}
\noindent where $V_{norm}$ the volume factor of volume $V$:
\begin{eqnarray}
V_{norm} = \log(V + 1)
\end{eqnarray}
We further compute the Pearson correlations between volume and $mAoC$ per field per year to quantify whether an increase in volume also comes with an increase of $mAoC$.

\noindent\textbf{Results.} Controlling for an increase in volume does not change the general trends of decrease in recent years. For example, NLP %
had a -10.4\% decline in VACA compared to -12.8\% in $mAoC$. Overall, fields have slightly smaller but yet marked decreases in volume-adjusted citation age.

We observe low Pearson correlations between volume and $mAoC$ for NLP (0.19), Medicine (0.21), CS (0.28), or Engineering (0.28). Other fields show marked correlations, such as psychology (0.49), math (0.71), and physics (0.72). When decomposing the correlations into consecutive decades, we see that CS shows small correlations in the past (0.18 from 1980 to 1990; -0.26 from 1990 to 2000) but recently has seen a wave of more volume and increasing citation age (0.62 from 2000 to 2010) followed by an anti-correlation (-0.54 from 2010 to 2020). More results are available in \Cref{ap:experiments}.

\noindent\textbf{Discussion.} In contrast to the concluding remarks of \citet{nguyen2024there}, we show there is a recency bias in multiple fields of study, even when controlling for paper volume and growth in annual papers. Several factors influence the dynamics of citations in different fields. Shifts in academic incentives play a crucial role; reviewers, institutions, and conferences can favor including more recent papers. This trend reflects evolving priorities within the academic community. Increasing pressure of the ``publish or perish'' principle in research has resulted in researchers splitting their work into minimum viable units that can be published. Thus, changes in citation amnesia (possibly caused by other factors) are further amplified by this change in behavior.
The rise of open-access movements and pre-print servers, which make papers immediately available, has likely also contributed to a trend of citing more recent works.\\[3pt]
\noindent\textbf{Q4.} \textit{How are different fields citing recent work from their own field against work from other fields?}

\noindent\textbf{Ans.} Academic fields tend to draw upon both their own historical literature as well as the work of other domains. Previous work has shown that intra-field citations grow over time for many fields \cite{wahle-etal-2023-cite}. However, it is unclear if different fields also cite more recent work from their own field compared to other fields. 

\begin{figure}[t]
    \centering
    \includegraphics[width=0.9\columnwidth]{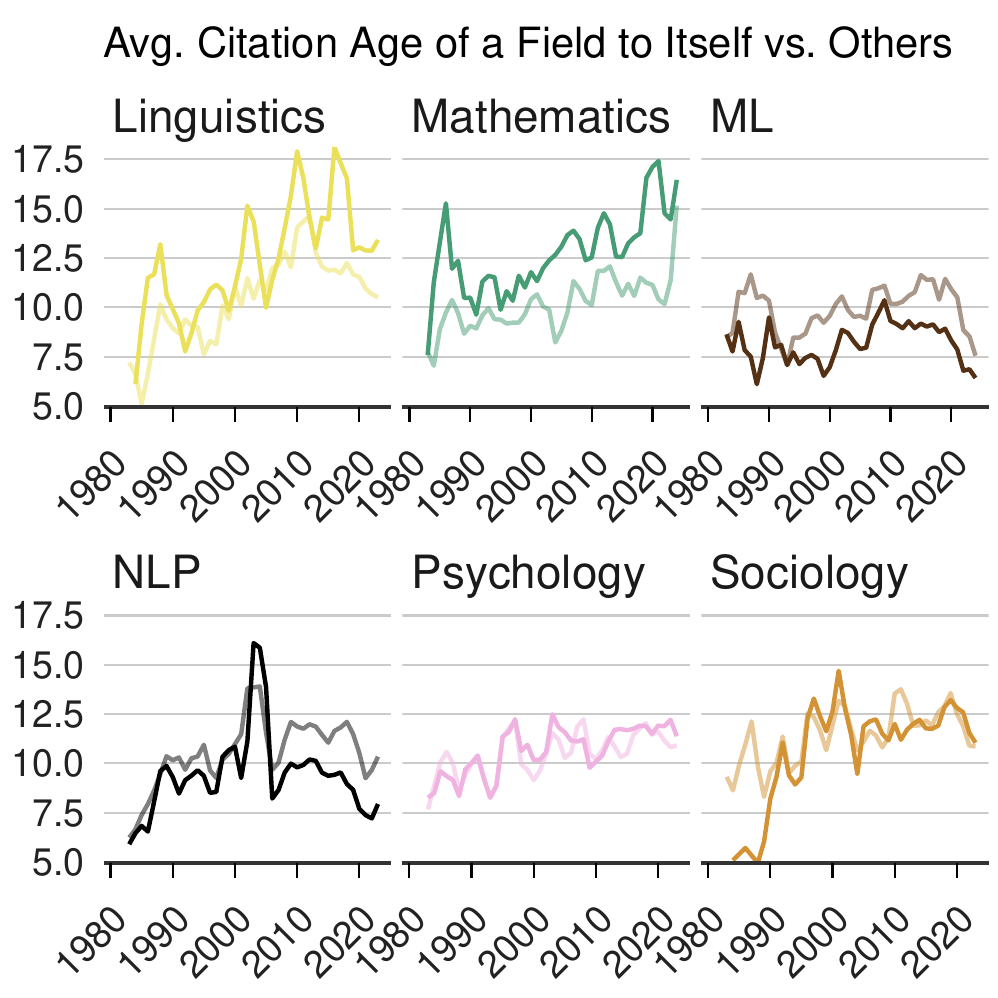}
    \caption{The $mAoC$ of NLP, ML, and the four most cited fields by NLP. Darker colors represent intra-field citations; lighter colors represent inter-field citations.}
    \label{fig:citation_age_to_same_field_vs_others}
\end{figure}

\noindent\textbf{Results.} The graph in \Cref{fig:citation_age_to_same_field_vs_others} shows the $mAoC$ for NLP, ML, and the four most cited fields by NLP (from 1980--2023). Observe that NLP and ML cite more recent work within their own field. The majority of fields cite slightly more recent work from their own field as opposed to work from other fields (14 of 20 fields). This is contrasted by fields such as linguistics and math, which tend to reference older works from within their own fields. The lines representing their intra-field citations have consistently been higher than those for citations from other fields over the years.

\noindent\textbf{Discussion.} Between fields, there is a notable variance in how they approach their own past academic work. 
Some fields, such as math and linguistics, show less tendency to not cite their own older works, which may be due to their long-standing history. 
In contrast, younger and evolving fields like NLP and ML are more focused on their own recent advancements, possibly due to the fast-paced nature of developments in these areas.\\[3pt]
\noindent\textbf{Q5.} \textit{How far back in time is NLP citing papers within CS compared to papers outside of CS?}

\noindent\textbf{Ans.} As \citet{bollmann-elliott-2020-forgetting,singh-etal-2023-forgotten} demonstrated, NLP has seen a shift towards citing more recent literature, particularly since around 2015. 
\citet{wahle-etal-2023-cite} further revealed that NLP papers predominantly cite works within CS. 
This raises an intriguing point about whether the NLP community focuses more on recent developments within CS, potentially at the expense of older, yet relevant, non-CS literature, or whether we cite more foundational non-CS work than CS.

\begin{figure}[t]
    \centering
    \includegraphics[width=0.85\columnwidth]{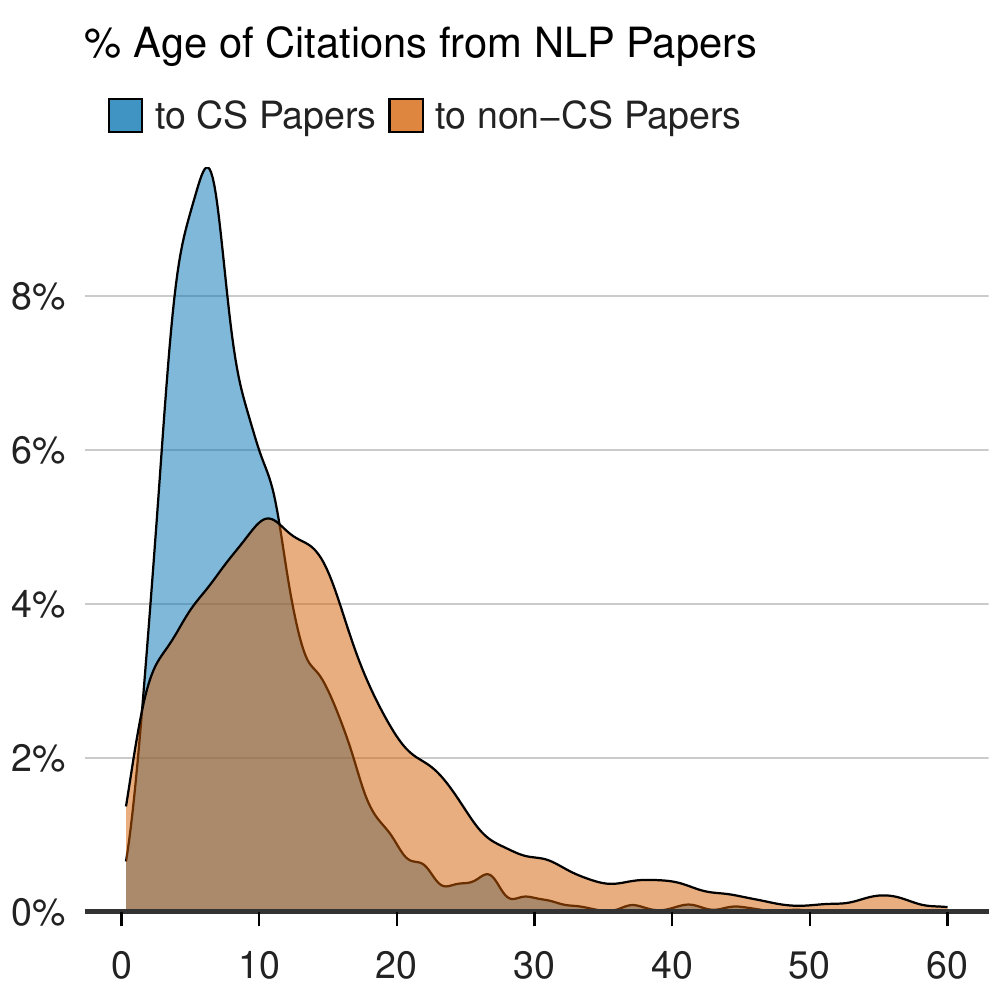}
    \caption{The percentage of $mAoC$ split for citations from NLP papers to CS papers and to non-CS papers.}
    \vspace*{4.5mm}\label{fig:citation_age_distr_cs_vs_non_cs}
\end{figure}

\noindent\textbf{Results.} \Cref{fig:citation_age_distr_cs_vs_non_cs} shows the distributions of $mAoC$ for NLP works citing CS papers and non-CS papers. There is a pronounced trend of citing recent CS papers, with most citations falling within the 4--10-year range with few papers being cited more than 30 years back in time. Non-CS papers tend to be cited much less within this timeframe; instead, they are more frequently cited when they are 10--20 years old with a marked proportion of papers being cited up to 40 years in the past.

\noindent\textbf{Discussion.} The `half-life' of ideas in CS appears to be shorter for NLP works, which could imply that older research is becoming less relevant faster than papers from outside of CS. 
This could be because NLP's (and other quantitative fields') research is concerned with more recent innovations (from CS) than from other fields or because disruptions occur faster in these technical fields than in others. 
Non-quantitative fields such as philosophy or sociology have a longer history and are arguably less iterative and more holistic.
These results raise questions about the sustainability of the innovation pace in these technical fields and whether it might lead to its continuous growth at a speed that may not allow for a thorough validation and understanding 
of past work. \\[3pt]
\noindent\textbf{Q6.} \textit{How much more in the past are we (NLP) citing various fields compared to the average of all other fields? And which other fields have cited NLP papers much more than this average in the past?}

\begin{figure}[t]
    \centering
    \includegraphics[width=0.85\columnwidth]{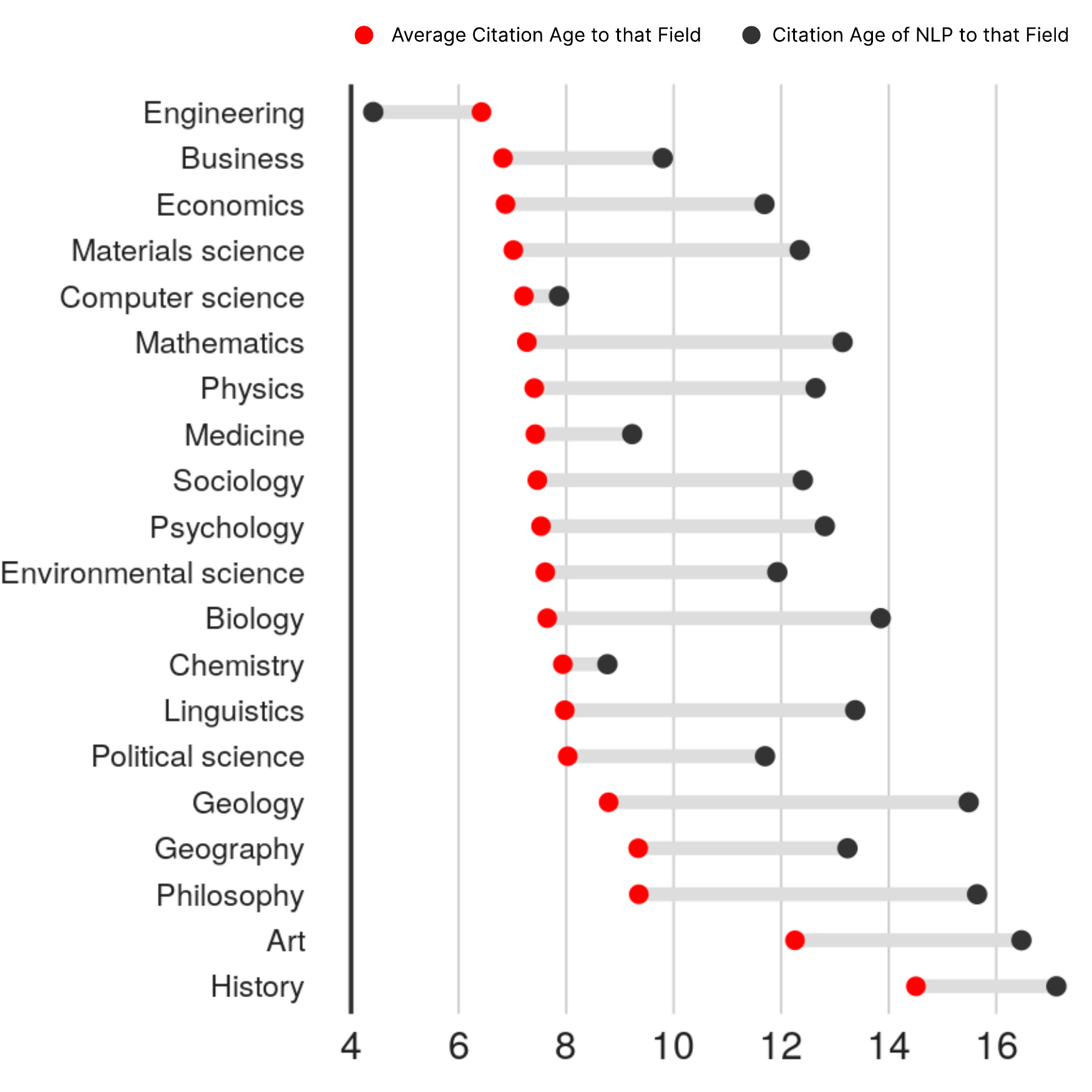}
    \caption{The macro-average $mAoC$ from each of the other fields to a target field (black). The $mAoC$ from NLP to a target field of study (red).}
\label{fig:citation_age_nlp_to_other_fields}
\end{figure}

\noindent\textbf{Ans.} As previous questions revealed, NLP has a particularly low $mAoC$ compared to other fields, and we are citing recent work mainly from within NLP. But are we citing works from specific other fields more or less in the past? For example, are we citing recent linguistic papers but old medicine papers? Or are we uniformly citing recent work across fields? To answer these questions, we measure $mAoC$ for papers in NLP citing papers in other fields and compare that to the micro-averaged $mAoC$ for any of the 20 fields citing that field (except intra-field citations, i.e., citations from papers in a field to papers in the same field).

\noindent\textbf{Results.} \Cref{fig:citation_age_nlp_to_other_fields} reveals marked differences between how far back in time NLP cites a field against the average field is citing that field. NLP cites recent engineering papers, with the $mAoC$ of NLP to that field being 4.1. CS papers are cited at a rate just less than a year below the average $mAoC$ across all fields, indicating a propensity within NLP to keep abreast of the latest CS research. NLP cites recent papers from fields like medicine and chemistry, whereas it draws on much older papers in math, linguistics, and physics, suggesting a reverence for foundational work in these areas. The average $mAoC$ for NLP to math and linguistics (the most highly cited fields of NLP \cite{wahle-etal-2023-cite}) stands out, showing that NLP research cites back to papers 13 years old on average.

\noindent\textbf{Discussion.} The recent citations from NLP to engineering may result from close ties and technological intertwining between the two fields. Some fields, like history, exhibit a high citation age due to the nature of their field (with inherent interest in the past).
NLP's tendency to cite older math, linguistics, and physics papers shows a long-term attribution to research that laid the groundwork for current NLP methods. What is yet unclear is whether we keep citing the same foundational works over time. \\[3pt]
\noindent\textbf{Q7.} \textit{Do the same papers remain highly cited or are different papers cited more in different periods?}

\begin{table}[t]
\centering
{\small
\begin{tabular}{cccc}
\toprule
\multirow{2}{*}{} & 90 -- 00 / & 00 -- 10 / & 10 -- 15 / \\
& 00 -- 10 & 10 -- 15 & 15 -- 20 \\
\midrule
80 -- 90 & -0.04 & -0.14 & 0.37 \\
90 -- 00 & - & {\space }0.16 & 0.46 \\
00 -- 10 & - & - & 0.28 \\
\bottomrule
\end{tabular}
}
\caption{We rank all papers in an epoch (e.g., 1980, 1990) by citations from two other epochs (e.g., 1990, 2000 and 2000, 2010). We compute Spearman correlations between both rankings: papers ranked by citations from the first range (e.g., 1990 -- 2000) and papers ranked by the second time range (e.g., 2000--2010).\vspace*{-4mm}}
\label{tab:shuffling}
\end{table}

\noindent\textbf{Ans.} On average, NLP papers cite fewer older papers and in different proportions from different fields (Q1, Q2, and Q6). 
What we do not know is which old works we are citing. Are there papers that are always cited? Or is there marked shuffling %
in the works being cited? %

To answer this question, we rank papers from a period A (say, 1990 to 2000) by citations received over two future and separate periods B (say 2000 to 2010) and C (say 2010 to 2020) and compute 
Spearman's rank correlation. %
We exclude papers with less than 10 total citations (as this experiment does not pertain to rarely cited papers).

\noindent\textbf{Results.} The Spearman correlation coefficients in \Cref{tab:shuffling} show weak to no correlation between the citation rankings of papers from the 1990s when compared to those in the 2000s and 2010s. %
However, there is a positive correlation (0.46) between the rankings of papers from 2000--2010 cited by 2010--2015 and 2015--2020. 
Manually examining the citation rankings of that epoch reveals that, generally, works that received a high number of citations in one epoch tended to keep their high citation count in subsequent epochs. 
This trend is more pronounced for works at the top positions, with less shuffling observed than in the lower-ranked papers.

\noindent\textbf{Discussion.} The considerable shuffling of citations between epochs shows that the factors influencing citation relevance are complex and multifaceted. Papers can fluctuate significantly in their citational importance over different periods. Such changes in citation frequency can be attributed to various reasons beyond forgetting. For example, some theories only become empirically testable with time as new data or methods become available. Instances like the LSTM network are examples where the original concepts were not immediately adopted but gained prominence later with advancements in computation and practical applicability \cite{hochreiter1997}.\\[3pt]
\noindent\textbf{Q8.} \textit{Is there an online tool that allows one to easily determine the citation age and fields cited by a paper (or a set of papers)?}

\noindent\textbf{Ans.} Yes. We have developed a freely accessible web-based tool to promote cognizance of temporal diversity in citations across different fields of study. Users can upload a paper's PDF, input an ACL Anthology or Semantic Scholar link (including author profiles or proceedings), and the system produces salient data and visualizations concerning the diversity of fields of the cited literature. More details on the demo are available in \Cref{ap:demo}.

\section{Concluding Remarks}

This work showed that many fields are experiencing a marked decline in citation age, which started between 2015 and 2019. 
NLP and ML show the strongest preference for citing recent works with a lower citation age than others.
Even when controlling for an increase in the volume of new papers, many fields cite recent work disproportional to the growth in papers.
We also show that NLP has a particular recency bias towards CS literature and cites recent work from engineering while citing older works from math and linguistics papers.

So, what does a falling rate of engagement with older research mean? Is this a natural development in the transition from a new and small field to a vibrant and large field of study? Or is this a symptom of an increasingly insular research culture that looks at its own past work at the expense of relevant outside work? It is not clear how this can be answered empirically, but we hope future work will address this.

The goal of our work is not to argue for either point but rather to help us reflect on our development as a field and scientific community. We must look harder at ourselves to ensure we are not developing bad practices.
This is especially important for NLP because of the 
the widespread deployment of its technologies into society at large. As many have argued, central to developing robust and ethical social systems is the engagement with a diverse array of literature spanning multiple disciplines and time periods. 
Further, as members of the scientific community, we can shape and direct the future of these trends by determining which ideas and works to engage with.\newline

\section*{Limitations}
This study examines 20 fields of study derived from the dataset used. Further, the borders between fields and whether and how much a focal scientific paper can be assigned to one or more fields are not fully defined and will always have overlapping regions. Each dataset also comes with its own biases. For example, \citet{bollmann-elliott-2020-forgetting,singh-etal-2023-forgotten} have focussed on the ACL Anthology (AA) for NLP papers, but there are many other papers in NLP outside of AA, \citet{nguyen2024there} have used preprints from arXiv, \citet{wahle-etal-2023-cite} have investigated Semantic Scholar (S2) which has indexed more biology and medicine papers than CS proportionally, while this study relied on OpenAlex which contains more CS papers. Although marked differences in datasets exist, this study has provided confirming results to \citet{bollmann-elliott-2020-forgetting,singh-etal-2023-forgotten,wahle-etal-2023-cite} and contrary results to \citet{nguyen2024there}.

This study looked at four decades of citational information across fields, which is a limited snapshot of scientific history, particularly for older, more foundational fields. Extending the investigation period could reveal whether similar increases and decreases have been observed in the past for different fields, what may have caused these changes, and how long they have existed. Also, this study has mainly looked at quantitative aspects of citation practices at a large-scale aggregate level while qualitative aspects could reveal the reasons behind why certain fields cite newer versus older literature. We are planning to extend this study in future work to provide more answers to these open questions.

\bibliography{anthology,custom}

\clearpage

\appendix

\section{Appendix}
\label{sec:appendix}

\subsection{Details on the Demo}
\label{ap:demo}
\begin{figure*}
    \centering
    \includegraphics[width=\textwidth]{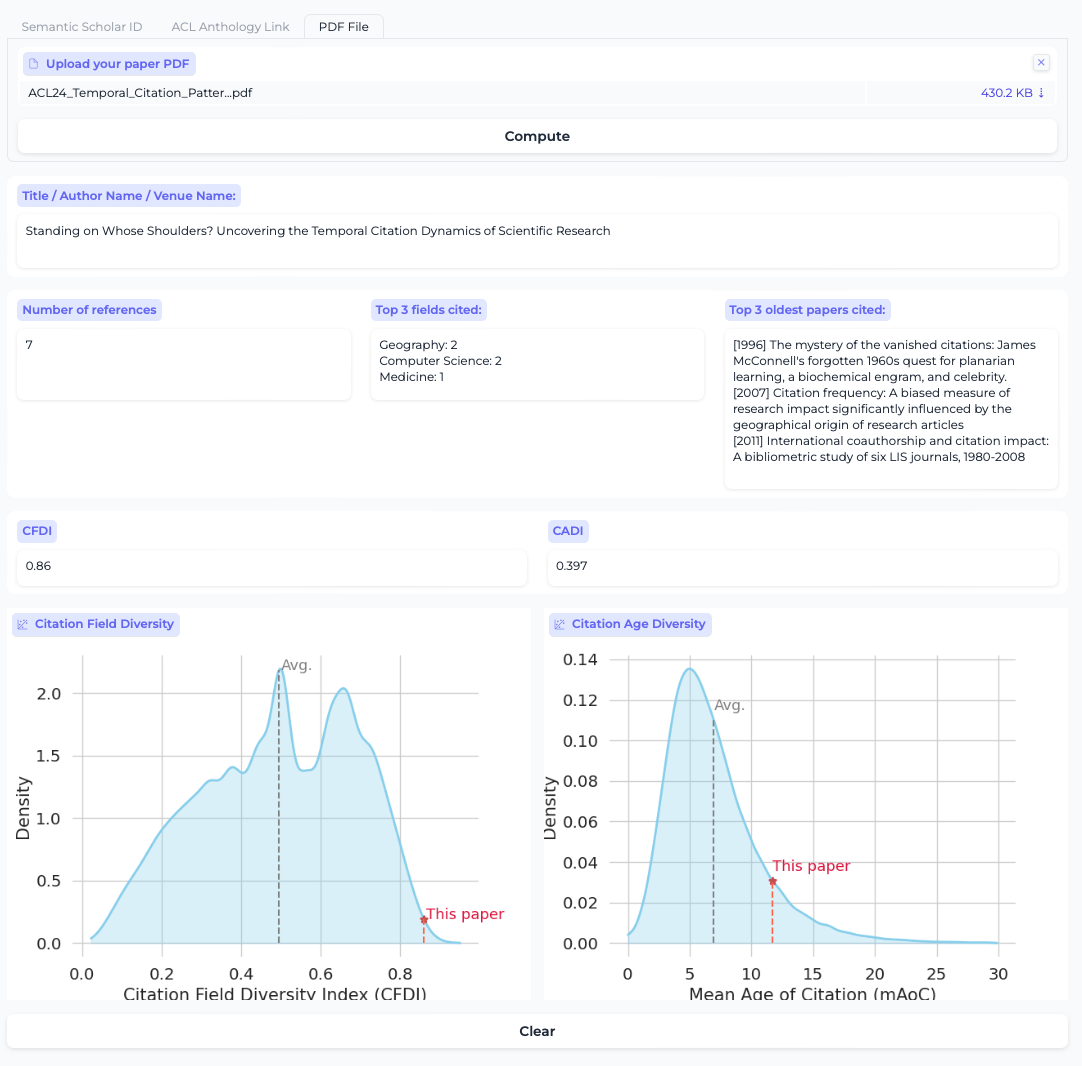}
    \caption{A free web demo to compute citation age and field diversity metrics for a paper, author, or proceeding given a PDF file, ACL Anthology link, or Semantic Scholar ID.}
    \label{fig:demo}
\end{figure*}

We have developed a freely accessible web-based tool to promote cognizance of temporal diversity in academic citations across different fields of study. Users can input any paper’s Semantic Scholar ID, ACL Anthology link, and PDF file, and our system yields salient data concerning the citation age and cross-field scope of the cited literature. One can also input author profiles or proceeding links from Semantic Scholar and the ACL Anthology. The interface visualizes the distribution of the citation age and citation field diversity for all NLP papers published until 2023, juxtaposing this with the citation age and citation field diversity of the paper inputted by the user. \Cref{fig:demo} shows an overview of the demo, which is available at

\begin{center}
    \url{https://huggingface.co/spaces/jpwahle/field-time-diversity}
\end{center}

\subsection{Supplementary Dataset Details}
\label{ap:data}
\Cref{tab:paper_counts_by_field} shows the number of papers per field, showing that CS, with 82.6m papers, is the largest field, followed by medicine, with 58.8m papers. Philosophy is the smallest one with 9.5m publications overall. AI has 14.5m papers, more than geology, environmental science, and philosophy.

\Cref{tab:field_group_mapping} shows the assignment of fields to the higher-level groups of \Cref{fig:perc_citations_to_older_works_field_groups} in Q2.

\begin{table}[t]
\resizebox{\columnwidth}{!}{
  \begin{tabular}{lrr}
  \toprule
  Field & Count & $\approx$1\% Sample \\
  \midrule
  Computer Science & 82,630,142 & 826,301 \\
  Medicine & 58,817,536 & 588,175 \\
  Biology & 43,427,071 & 434,271 \\
  Physics & 40,120,421 & 401,204 \\
  Political science & 34,186,267 & 341,863 \\
  Chemistry & 34,005,729 & 340,057 \\
  Engineering & 31,181,385 & 311,814 \\
  Philosophy & 30,885,218 & 308,852 \\
  Mathematics & 28,048,330 & 280,483 \\
  Psychology & 25,187,604 & 251,876 \\
  Materials Science & 21,913,736 & 219,137 \\
  Art & 21,010,953 & 210,110 \\
  Geography & 19,189,950 & 191,900 \\
  Business & 18,518,709 & 185,187 \\
  Sociology & 17,345,207 & 173,452 \\
  Economics & 16,727,938 & 167,279 \\
  Artificial Intelligence$^*$ & 14,456,606 & 144,566 \\
  Geology & 13,380,595 & 133,806 \\
  History & 12,488,890 & 124,889 \\
  Environmental Science & 11,482,177 & 114,822 \\
  Philosophy & 9,481,905 & 94,819 \\
  Machine Learning$^*$ & 3,663,369 & 36,634 \\
  Natural Language Processing$^*$ & 964,937 & 30,000 \\
  \bottomrule
  \end{tabular}
}
\caption{Number of papers per field for all 23 fields with ~1\% sample size. $^*$These are second-level subfields of the others, not fields of their own.} \label{tab:paper_counts_by_field}
\end{table}

\begin{table*}[ht]
{\small
\centering
\resizebox{\textwidth}{!}{
    \begin{tabular}{ll}
    \toprule
    \textbf{Group}      & \textbf{Fields}                                                                                         \\ \midrule
    Social Sciences     & Political Science, Psychology, Sociology, Economics, Geography, Business                                \\ 
    Natural Sciences    & Biology, Physics, Chemistry, Environmental Science, Medicine, Geology, Materials Science               \\ 
    Formal Sciences     & Computer Science, Mathematics, AI, ML, NLP                                                              \\ 
    Humanities          & Art, History, Philosophy                                                                                \\ 
    \bottomrule
    \end{tabular}
}
}
\caption{Mapping of fields to four higher-level groups of \Cref{fig:perc_citations_to_older_works_field_groups}.}
\label{tab:field_group_mapping}
\end{table*}

\subsection{Supplemental Experimental Results} \label{ap:experiments}

We provide additional results on Q3 and the Pearson correlation experiments between volume and $mAoC$ in \Cref{tab:pearson_correlations}.

\begin{table*}[t]
\centering
\begin{tabular}{lrrrrr}
\toprule
 & Overall & 1980-1990 & 1990-2000 & 2000-2010 & 2010-2020 \\
\midrule
NLP & 0.19 & 0.12 & -0.29 & 0.65 & -0.47 \\
ML & 0.29 & 0.15 & -0.22 & 0.75 & -0.50 \\
Art & 0.19 & -0.12 & -0.16 & 0.59 & 0.08 \\
Biology & 0.50 & -0.04 & -0.03 & 0.45 & 0.16 \\
Business & 0.47 & 0.34 & -0.22 & 0.59 & 0.47 \\
Chemistry & 0.64 & 0.29 & 0.49 & 0.45 & -0.25 \\
Computer science & 0.28 & 0.18 & -0.26 & 0.62 & -0.55 \\
Economics & 0.16 & 0.53 & -0.03 & 0.26 & 0.38 \\
Engineering & 0.28 & 0.27 & 0.44 & -0.29 & 0.28 \\
Environmental science & 0.45 & 0.07 & -0.24 & 0.11 & 0.33 \\
Geography & 0.16 & -0.25 & 0.37 & -0.21 & 0.10 \\
Geology & 0.23 & 0.51 & 0.43 & 0.03 & 0.44 \\
History & -0.03 & -0.14 & 0.37 & -0.18 & 0.63 \\
Materials science & 0.48 & -0.17 & 0.43 & 0.33 & -0.24 \\
Mathematics & 0.71 & 0.76 & 0.42 & 0.67 & 0.29 \\
Medicine & 0.21 & 0.39 & 0.09 & 0.12 & -0.35 \\
Philosophy & 0.17 & -0.02 & 0.14 & 0.24 & -0.08 \\
Physics & 0.72 & 0.16 & 0.25 & 0.47 & 0.32 \\
Political science & 0.12 & 0.59 & -0.08 & 0.14 & 0.34 \\
Psychology & 0.49 & -0.23 & 0.37 & -0.08 & 0.76 \\
Sociology & 0.08 & -0.26 & -0.17 & 0.33 & 0.43 \\
\bottomrule
\end{tabular}
\caption{Pearson correlation between the volume of papers and $mAoC$ across different time ranges.}
\label{tab:pearson_correlations}
\end{table*}

\subsection{Additional Research Questions}

We also extended our analysis to the connection between citation age and field diversity as well as citation age for different institutions with two additional questions answered in the following.

\noindent\textbf{AQ1.} \textit{Does citing papers from different fields correlate with citing papers from different periods in time? In other words, does the diversity of citing a broad set of fields correlate with the diversity of citation ages?}

\begin{table*}[t]
\centering
\resizebox{\textwidth}{!}{
    \begin{tabular}{lrrrr}
    \toprule
    Field & Inc. / Out. CAD & Inc. / Out. CFD & Inc. CFD / CAD & Out. CFD / CAD \\
    \midrule
    NLP & -0.03 & 0.48 & 0.44 & 0.44 \\
    ML & -0.07 & 0.42 & 0.48 & 0.39 \\
    Art & -0.10 & 0.27 & 0.48 & 0.48 \\
    Biology & -0.07 & 0.44 & 0.35 & 0.32 \\
    Business & -0.10 & 0.26 & 0.52 & 0.48 \\
    Chemistry & -0.15 & 0.38 & 0.40 & 0.36 \\
    Computer science & -0.09 & 0.41 & 0.50 & 0.42 \\
    Economics & -0.09 & 0.35 & 0.48 & 0.41 \\
    Engineering & -0.07 & 0.36 & 0.48 & 0.42 \\
    Environmental science & -0.13 & 0.26 & 0.52 & 0.41 \\
    Geography & -0.10 & 0.34 & 0.53 & 0.43 \\
    Geology & -0.09 & 0.42 & 0.46 & 0.38 \\
    History & -0.12 & 0.35 & 0.52 & 0.53 \\
    Linguistics & -0.11 & 0.35 & 0.45 & 0.38 \\
    Materials science & -0.12 & 0.42 & 0.43 & 0.37 \\
    Mathematics & -0.10 & 0.41 & 0.47 & 0.38 \\
    Medicine & -0.10 & 0.50 & 0.43 & 0.36 \\
    Philosophy & -0.12 & 0.32 & 0.51 & 0.45 \\
    Physics & -0.13 & 0.39 & 0.44 & 0.35 \\
    Political science & -0.06 & 0.24 & 0.53 & 0.50 \\
    Psychology & -0.09 & 0.39 & 0.49 & 0.34 \\
    Sociology & -0.09 & 0.34 & 0.48 & 0.41 \\
    \bottomrule
    \end{tabular}
}
\caption{Spearman correlation between yearly metrics of CFD and CAD across various fields. We calculate Spearman correlation for each metric, \( x \) and \( y \), for each year, where \( x \) and \( y \) are incoming and outgoing CFD and CAD. All results are statistically significant with \( p < 0.05 \).}
\label{tab:spearman_correlations}
\end{table*}

\noindent\textbf{Ans.} Previous studies have introduced two metrics to capture how diversely papers cite across time \cite{singh-etal-2023-forgotten,nguyen2024there}, the citation age diversity (CAD), and how diversely they cite across fields, the citation field diversity (CFD). CAD applies the 1 - Gini index to the AoCs of a paper, while CFD applies Gini-Simpson to the counts of citations per field. CAD scores close to one means the paper cites other papers equally across time, while a value close to zero means all citations are concentrated in one year. CFD scores close to one means all citations are equally distributed across fields, while a value of zero means all citations are concentrated in a single field.

As this and previous studies have underlined, NLP's tendency to cite papers diverse across the past \cite{bollmann-elliott-2020-forgetting,singh-etal-2023-forgotten,nguyen2024there}, and to cite papers from a diverse set of fields \cite{wahle-etal-2023-cite} are decreasing. Q4 also showed that often, fields have more recent intra-field citations. Using CAD and CFD, we can quantify whether citing less work from different fields (partially) explains a decrease in citation age as well. In other words, are these two variables, CFD and CAD, correlated? And how much do incoming CAD and outgoing CAD correlate, and the same for incoming and outgoing CFD? To answer these questions, we calculate the Spearman correlation between CAD and CFD for both incoming and outgoing citations.

We expect how far back in time we cite is not linked to how far in the future we get cited, i.e., incoming CAD and outgoing CAD are not correlated. How much we draw from other fields could be linked to how much other fields draw from us, meaning there could be a correlation between incoming and outgoing CFD. Also, citing back in time and various fields (incoming CFD and CAD as well as outgoing CFD and CAD) could be linked because of the intra-/inter-field citation age discrepancy seen in Q5 and Q6.

\textbf{Results.} The results in \Cref{tab:spearman_correlations} (first column) indicate that, as expected, there is no correlation between the age of citations a field receives (incoming) and the age of references it cites (outgoing) across fields. However, there is a slight positive correlation between the diversity of fields (second column) a paper cites (outgoing) and the diversity of fields from which it receives citations (incoming). Notably, NLP, ML, and medicine show moderate correlations between incoming and outgoing CFD --- fields with low mean $mAoC$. 

When looking at correlations between incoming CFD and CAD (third column) and outgoing CFD and CAD (fourth column), similar positive correlations can be observed.
This suggests that papers that draw from a wide range of fields also tend to attract citations from a diverse range of time.

\textbf{Discussion.} Different fields also have different temporal citation patterns; therefore, citing widely across fields can also lead to more diversity in citation ages. 
Also, integrating ideas from a diverse set of fields can lead to wider relevance across different fields (as opposed to a single field) and, thus, a broader and longer citation base.

\noindent\textbf{AQ2.} \textit{What is the citation age of various companies, educational institutions, and governments?}

\noindent\textbf{Ans.} In evaluating the average age of citations across various institutions, we computed the $mAoC$ for publications affiliated with sectors of education, government, and the corporate world, which we manually selected based on data coverage and volume of research output.

\begin{figure}[t]
    \centering
    \includegraphics[width=0.9\columnwidth]{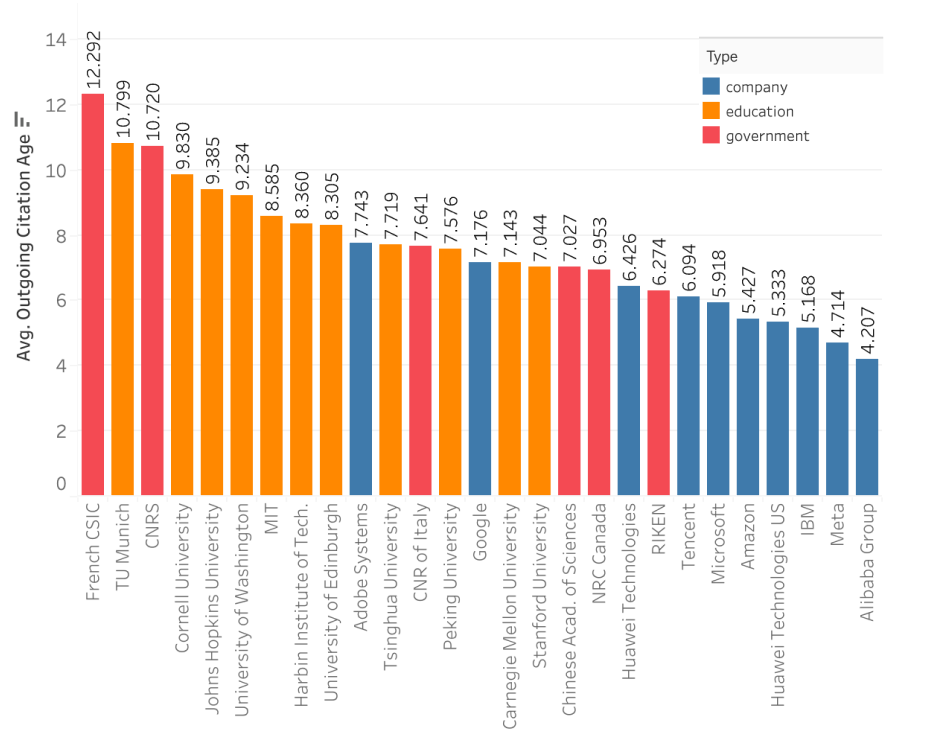}
    \caption{The $mAoC$ for different institutions.}
    \label{fig:citation_age_institutions}
\end{figure}

\noindent\textbf{Results.} \Cref{fig:citation_age_institutions} shows entities such as French CNRS, TU Munich, and Cornell University among educational institutions; Google, IBM, and Microsoft represent companies; and the NIH and NRC Canada for government institutions. Educational institutions tend to cite older works, with the average citation age reaching up to approximately 12 years, suggesting a scholarly inclination towards classical and foundational literature. Corporate citations are more contemporary, averaging between 5 and 7 years. The citation age of government institutions, on average, lies in between education and industry.

\noindent\textbf{Discussion.} Differences in citation age indicate each sector's underlying motivations and research ethos. Corporations may focus on cutting-edge studies to foster innovation, whereas academic institutions often incorporate a mix of historical and modern literature to support education and research. The upfront cost of foundational research with no short- or medium-term financial return can threaten a company's success. At the same time, institutions and government can rely on public funding to explore foundational questions.

\subsection{AI Usage Card}

We report how we used AI assistants such as ChatGPT and Gemini for this work in the following standardized card according to \citet{wahle2023ai}.

\onecolumn

{\sffamily
    \centering
    \tcbset{colback=white!10!white}
    \begin{tcolorbox}[
        title={\large \textbf{AI Usage Card} \hfill \makebox{\qrcode[height=1cm]{https://ai-cards.org}}},
        breakable,
        boxrule=0.7pt,
        width=.8\paperwidth,
        center,
        skin=bicolor,
        before lower={\footnotesize{AI Usage Card v1.0 \hfill \url{https://ai-cards.org} \hfill PDF | BibTeX}},
        segmentation empty,
        halign lower=center,
        collower=white,
        colbacklower=tcbcolframe]
            
        \footnotesize{
            \begin{longtable}{p{.15\paperwidth} p{.275\paperwidth} p{.275\paperwidth}}
              {\color{LightBlue} \MakeUppercase{Correspondence(s)}} \newline Jan Philip Wahle 
              & {\color{LightBlue} \MakeUppercase{Contact(s)}} \newline wahle@uni-goettingen.de %
              & {\color{LightBlue} \MakeUppercase{Affiliation(s)}} \newline University of Göttingen %
              \\\\
              & {\color{LightBlue} \MakeUppercase{Project Name}} \newline Citation Amnesia: On The Recency Bias of NLP and Other Academic Fields
              & {\color{LightBlue} \MakeUppercase{Key Application(s)}} \newline Citation Analysis, Scientometrics, NLP
              \\\\
              {\color{LightBlue} \MakeUppercase{Model(s)}} \newline ChatGPT \newline Gemini
              & {\color{LightBlue} \MakeUppercase{Date(s) Used}} \newline 2023-10-01 \newline 2024-01-01
              & {\color{LightBlue} \MakeUppercase{Version(s)}} \newline 4.0 \newline Ultra\\\\
              \cmidrule{2-3}\\
      
              {\color{LightBlue} \MakeUppercase{Ideation}} \newline    
              & {\color{LightBlue} \MakeUppercase{Generating ideas, outlines, and workflows}} \newline Not used  
              & {\color{LightBlue} \MakeUppercase{Improving existing ideas}} \newline Not used  \\\\
              & {\color{LightBlue} \MakeUppercase{Finding gaps or compare aspects of ideas}} \newline Not used \\\\
              
              {\color{LightBlue} \MakeUppercase{Literature Review}} \newline    
              & {\color{LightBlue} \MakeUppercase{Finding literature}} \newline Not used
              & {\color{LightBlue} \MakeUppercase{Finding examples from known literature}} \newline Not used \\\\
              & {\color{LightBlue} \MakeUppercase{Adding additional literature for existing statements and facts}} \newline Not used
              & {\color{LightBlue} \MakeUppercase{Comparing literature}} \newline Not used \\\\
              \cmidrule{2-3}\\
      
              {\color{LightBlue} \MakeUppercase{Methodology}} \newline    
              & {\color{LightBlue} \MakeUppercase{Proposing new solutions to problems}} \newline Not used
              & {\color{LightBlue} \MakeUppercase{Finding iterative optimizations}} \newline Not used \\\\
              & {\color{LightBlue} \MakeUppercase{Comparing related solutions}} \newline Not used \\\\
              
              {\color{LightBlue} \MakeUppercase{Experiments}} \newline    
              & {\color{LightBlue} \MakeUppercase{Designing new experiments}} \newline Not used
              & {\color{LightBlue} \MakeUppercase{Editing existing experiments}} \newline Not used \\\\
              & {\color{LightBlue} \MakeUppercase{Finding, comparing, and aggregating results}} \newline Not used \\\\
              \cmidrule{2-3}\\
      
              {\color{LightBlue} \MakeUppercase{Writing}} \newline ChatGPT Gemini  
              & {\color{LightBlue} \MakeUppercase{Generating new text based on instructions}} \newline Used
              & {\color{LightBlue} \MakeUppercase{Assisting in improving own content}} \newline Used \\\\
              & {\color{LightBlue} \MakeUppercase{Paraphrasing related work}} \newline Used 
              & {\color{LightBlue} \MakeUppercase{Putting other works in perspective}} \newline Not used \\\\
              
              {\color{LightBlue} \MakeUppercase{Presentation}} \newline    
              & {\color{LightBlue} \MakeUppercase{Generating new artifacts}} \newline Not used
              & {\color{LightBlue} \MakeUppercase{Improving the aesthetics of artifacts}} \newline Not used \\\\
              & {\color{LightBlue} \MakeUppercase{Finding relations between own or related artifacts}} \newline Not used \\\\
              \cmidrule{2-3}\\
              {\color{LightBlue} \MakeUppercase{Coding}} \newline ChatGPT   
              & {\color{LightBlue} \MakeUppercase{Generating new code based on descriptions or existing code}} \newline Used
              & {\color{LightBlue} \MakeUppercase{Refactoring and optimizing existing code}} \newline Used \\\\
              & {\color{LightBlue} \MakeUppercase{Comparing aspects of existing code}} \newline Not used \\\\
              
              {\color{LightBlue} \MakeUppercase{Data}} \newline    
              & {\color{LightBlue} \MakeUppercase{Suggesting new sources for data collection}} \newline Not used 
              & {\color{LightBlue} \MakeUppercase{Cleaning, normalizing, or standardizing data}} \newline Not used  \\\\
              & {\color{LightBlue} \MakeUppercase{Finding relations between data and collection methods}} \newline Not used  \\\\
              \cmidrule{2-3}\\
      
              {\color{LightBlue} \MakeUppercase{Ethics}} \newline    
              & {\color{LightBlue} \MakeUppercase{What are the implications of using AI for this project?}} \newline Generating code and improving the clearity of writing the paper has improved the efficacy of performing this scientific work.
              & {\color{LightBlue} \MakeUppercase{What steps are we taking to mitigate errors of AI for this project?}} \newline We manually fact-checked generated texts and inspected source code for potential generated bugs. \\\\
              & {\color{LightBlue} \MakeUppercase{What steps are we taking to minimize the chance of harm or inappropriate use of AI for this project?}} \newline We did not include text suggestions that had any chance of impacting marginalized groups.
              & {\color{LightBlue} \MakeUppercase{The corresponding authors verify and agree with the modifications or generations of their  used AI-generated content}} \newline Yes \\
      
            \end{longtable}
        }
        \tcblower
    \end{tcolorbox}
}

\cleardoublepage
\onecolumn
\hypertarget{annotation}{}
\citationtitle

\onlineversion{https://arxiv.org/abs/2402.12046}
\begin{bibtexannotation}
@inproceedings{wahle-etal-2025-citationamnesia,
	title        = {Citation Amnesia: On The Recency Bias of Natural Language Processing and Other Academic Fields},
	author       = {Wahle, Jan Philip and Ruas, Terry and Abdalla, Mohamed and Gipp, Bela and Mohammad, Saif},
	year         = 2025,
	booktitle    = {Proceedings of the International Conference on Computational Linguistics (COLING 2025)}
}
\end{bibtexannotation}

\end{document}